\documentclass[journal,onecolumn]{IEEEtran}

\usepackage{PRIMEarxiv}

\usepackage[utf8]{inputenc} 
\usepackage[T1]{fontenc}    
\usepackage{hyperref}       
\usepackage{url}            
\usepackage{booktabs}       
\usepackage{amsfonts}       
\usepackage{amsmath}
\usepackage{nicefrac}       
\usepackage{microtype}      
\usepackage{lipsum}
\usepackage{fancyhdr}       
\usepackage{graphicx}       
\usepackage{caption}
\usepackage{subcaption}
\usepackage{chemformula}
\usepackage{float}
\usepackage{lastpage}

\usepackage[numbers]{natbib}

\usepackage{fancyhdr}
\graphicspath{{media/}}     

\title{
Pushing the Limits of Quantum Computing for Simulating PFAS Chemistry
}

\author{
    Emil Dimitrov, Goar S{\'a}nchez-Sanz, James Nelson, Lee O'Riordan, Myles Doyle, \\
    \textbf{Se{\'a}n Courtney}, \textbf{Venkatesh Kannan}{\thanks{Corresponding author.}} \\
    Irish Centre for High-End Computing (ICHEC) \\
    \texttt{\{emil.dimitrov, venkatesh.kannan\}@ichec.ie}, \texttt{goar.sanchezsanz@manchester.ac.uk}, \\
    \texttt{janelson@tcd.ie}, \texttt{loriordan@gmail.com}, \texttt{mylesdoyle7@gmail.com}, \texttt{seancourtney51@gmail.com} \\
  \And
    Hassan Naseri, Alberto Garc{\'i}a Garc{\'i}a, James Tricker , Marisa Faraggi\\
    Accenture \\
    \texttt{\{hassan.naseri, a.b.garcia.garcia, james.tricker, marisa.faraggi\}@accenture.com} \\
  \And
    Joshua Goings, Luning Zhao \\
    IonQ \\
    \texttt{\{goings, zhao\}@ionq.co}
}

\begin{document}
\bstctlcite{IEEEexample:BSTcontrol}
\maketitle

\begin{abstract}
Accurate and scalable methods for computational quantum chemistry can accelerate research and development in many fields, ranging from drug discovery to advanced material design. 
Solving the electronic Schr\"odinger equation is the core problem of computational chemistry. 
However, the combinatorial complexity of this problem makes it  intractable to find exact solutions, except for very small systems. 
The idea of quantum computing originated from this computational challenge in simulating quantum-mechanics. 
We propose an end-to-end quantum chemistry pipeline based on the variational quantum eigensolver (VQE) algorithm and integrated with both HPC-based simulators and a trapped-ion quantum computer. 
 Our platform orchestrates hundreds of simulation jobs on compute resources to efficiently complete a set of \emph{ab initio} chemistry experiments with a wide range of parameterization. 
%

Per- and poly-fluoroalkyl substances (PFAS) are a large family of human-made chemicals that pose a major environmental and health issue globally.
Our simulations includes breaking a Carbon-Fluorine bond in trifluoroacetic acid (TFA), a common PFAS chemical.
This is a common pathway towards destruction and removal of PFAS.
Molecules are modeled on both a quantum simulator and a trapped-ion quantum computer, specifically IonQ Aria. 
Using basic error mitigation techniques, the 11-qubit TFA model (56 entangling gates) on IonQ Aria yields near-quantitative results with milli-Hartree accuracy.
Our novel results show the current state and future projections for quantum computing in solving the electronic structure problem, push the boundaries for the VQE algorithm and quantum computers, and facilitates development of quantum chemistry workflows.  
\end{abstract}



\newpage
\section{Introduction}
\label{sec:introduction_and_background}

Per- and poly-fluoroalkyl substances (PFAS) are a large family of human-made molecules that are extensively used for their water and oil repellent, heat resistant, and non-stick properties in areas such as clothing, adhesives, fire retardants, food packaging, cooking utensils, and paints.
PFAS are highly resistant to bio-degradation, and have high mobility in water and soils. 
PFAS can contaminate drinking water \cite{EUSPA2016} and bio-accumulate in wildlife.
PFAS exposure has detrimental health impacts on humans including cancer, thyroid diseases, hormonal changes, and reduced vaccine response, among others \cite{PFAS2016,Looker2013}. 
Therefore, remediation of PFAS from the ecosystem is crucial and has been widely studied both from experimental and computational/theoretical points of view \cite{DOMBROWSKI2018}.

Several techniques have been developed to physically remove PFAS from drinking water, for example, absorption of PFAS in active carbon, water filtration, or using ion exchange techniques\cite{RAHMAN2014,Kumarasamy2020,XIAO2017}. However, once the molecules are isolated, they must be still degraded and destroyed. 

\begin{figure}[htb!]
\centering
\includegraphics[scale=0.4]{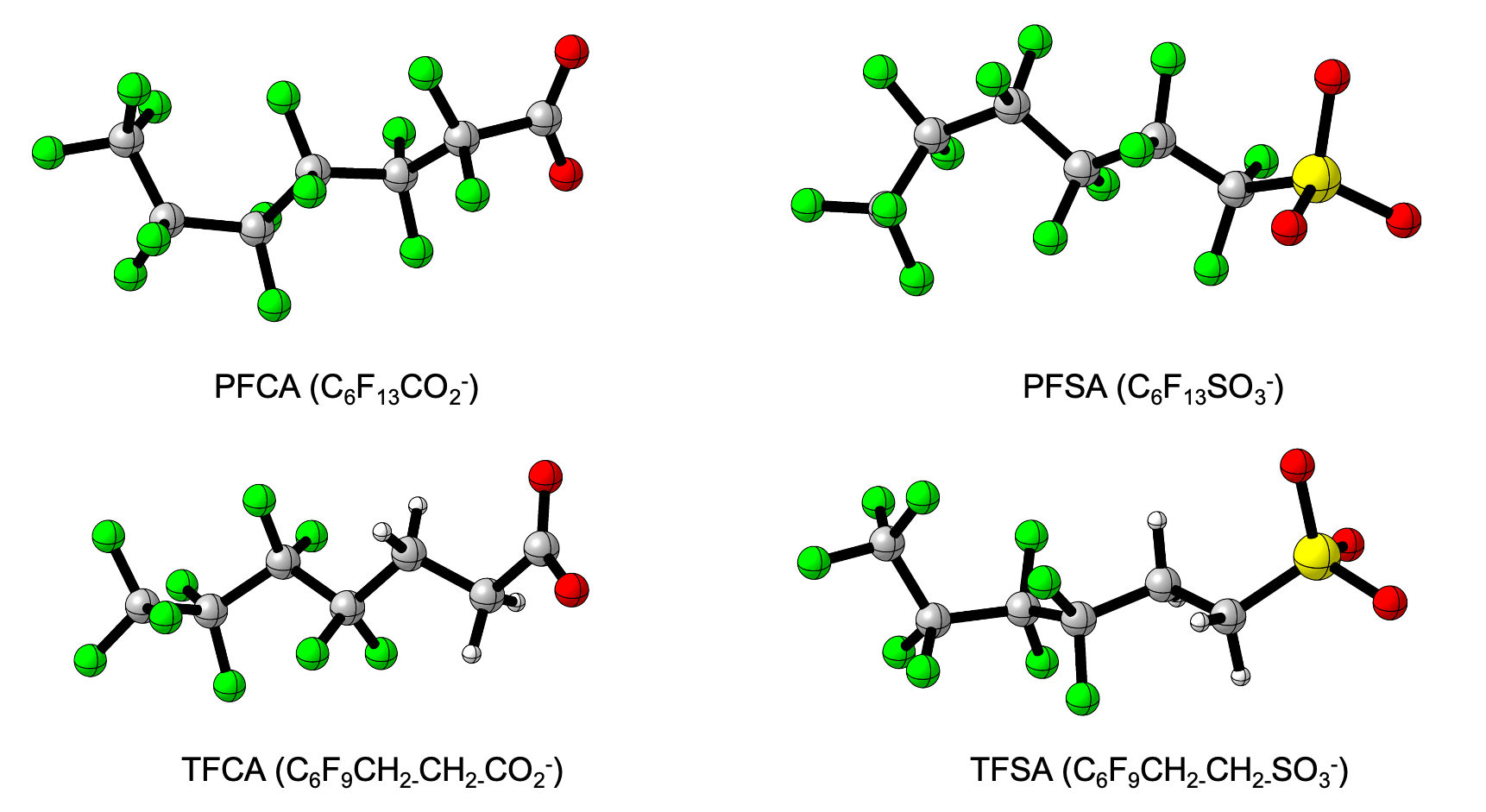}
\caption{Representative structures of PFAS molecules that are common in the environment: PFCA (perfluorocarbonylic acid), PFSA (perfluorosulfonic acid), TFCA (fluorotelomer carboxylic acid) and TFSA (trifluoromethanesulfonic acid) with n=6; n indicating the variable length of C$_n$F$_{2n+1}$}
\label{fig:representative_structures_for_qpfas}
\end{figure}

PFAS family presents an unusual thermal stability, mainly due to the large electro-negativity of fluorine, which forms strong C-F bonds. 
These bonds result in high surface activity and thermal resistance, as well as resistance to many forms of physical, chemical, and biological degradation. 
From chemical point of view, PFAS degradation presents a huge technical challenge, which depends on the type of PFAS, the length of the chains, and additional substituents within the compound. 
As an example, Figure~\ref{fig:representative_structures_for_qpfas} illustrates perfluorocarbonylic acids, PFCAs (C$_n$F$_{2n+1}$-COO$^-$), and perfluorosulfonic acids, PFSAs (C$_n$F$_{2n+1}$-SO$_3^-$) which are less stable than their telomeric counterparts, TFCA (C$_n$F$_{2n+1}$-CH$_2$-CH$_2$-COO$^-$) and TFSA (C$_n$F$_{2n+1}$-CH$_2$-CH$_2$-SO$_3^-$) \cite{BENTHEL2019}.

In the last few years, there has been significant development in PFAS destruction technologies. 
Examples include wet oxidation using heat activated oxidants \cite{BRUTON2017,PARK2016} and ion exchange using plasma reactors \cite{SALEEM2020123031,SINGH202} among other methods. 
Most of these techniques involve the use of strong oxidative species (mostly a variety of radicals) and/or reductive species (e.g. hydrated electron e$^{-}_{aq}$ produced by using UV light) \cite{liu2021}. 
The main objective of such techniques is to degrade PFAS molecules into smaller and treatable units. 
From the structural perspective, two main strategies can be considered: defluorination (dissociation of C-F bonds) and degradation of the backbone chain into smaller parts. However, these methods are even more challenging since C-F is among one of the strongest bonds in chemistry. Therefore, an attractive and potentially useful challenge is to computationally investigate PFAS degradation through defluorination and breaking C-F bonds.

For this computational chemistry challenge, there is a large number of possible experiments involving the dissociation of a single fluorine atom depending on the size of the PFAS, the number of C-F bonds and their positions. 
Computational coupled-cluster-based methods such as CCSD or CCSD(T) are preferable to achieve chemical accuracy, since these methods have been established as reliable and accurate for the computation of chemical energies.
CCSD(T) is often referred to as the ``gold-standard'' of electronic structure. 
However, these methods scale at $N^6$ and $N^7$ with the number of basis functions $N$, leading to high computational cost. 
As the number of basis functions grows with the number of atoms in a molecule, the computational expense rapidly grows with atom count. 
Hence, applying these high-accuracy methods to large molecules such as PFAS is impractical.

In this project we focus on the molecular simulation and analysis of the electronic and chemical properties of PFAS-like molecular structures. 
Currently, high-performance computing (HPC) models and tools are used for simulation and analysis of molecular structures and their electronic/chemical properties. However, current HPC simulations are computationally expensive and not particularly scalable, which limits the accuracy, size, and number of compounds subject to study. Approximations are necessary to reduce the computational cost and also to map molecular and electronic properties, which are inherently quantum, for execution on classical computation platforms.

With the advent of quantum computing hardware and algorithms, there has been interest in applying quantum computational methods to the simulation and analysis of the electronic structure systems. 
The scope of this project is to develop a high-performance hybrid quantum computational workflow with the aim of performing scalable and repeatable experiments using Variational Quantum Eigensolver (VQE) -based methods\cite{Peruzzo2014-xz, OMalley2016-ph, Colless2018-hi, McCaskey2019-gk, Nam2020-ct, Kandala2017-kr, Kandala2019-rz, Gao2019-pp, Rice2021-mn, Gao2021-zt, Zhao2023-um, Goings2023-qr}. 
We demonstrate the development of an end-to-end workflow to run hybrid high-performance quantum computational chemistry experiments with a wide range of choices in the parameter space (e.g. molecule types, active spaces, qubit reduction methods, chemistry parameters, and VQE-related parameters).  
This workflow can interface with quantum computing software simulators as well as quantum computing hardware. 
The pipeline is used to compute the energies in a bond-breaking experiment for PFAS-related molecules both in simulations and on a real quantum hardware.

The rest of this manuscript is organized as follows.
Section 2 describes the implementation of the workflow in details.
Section 3 contains the results, which are divided into two subsections: the results obtained on a quantum simulator, and the results from a real quantum computer. 
Finally, we summarize the work in Section 4.    

\section{Implementation}
\label{sec:workflow_and_implementation}
As outlined in the introduction, the dissociation energy of C-F bonds is studied in this paper.
Breaking a C-F bond is a fundamental step in the degradation process of PFAS.
To simulate bond dissociation, a target molecule is selected and a dictionary of geometries is generated.
Then, the position of one of the atoms, usually the outermost (e.g. fluorine atom in TFA), is shifted in steps to simulate bond breaking and evaluate the energy surface at different distances.
The electronic structure problem for these geometries are first solved at the Hartree-Fock self-consistent field level. 
From these results the one- and two-body electronic tensors are obtained to form the fermionic Hamiltonian, which are then solved using VQE to obtain an estimate of the ground state energy.
The fermionic Hamiltonian must be mapped to qubit operations to be executed on a quantum hardware. 
This transformation can be efficiently performed classically using, e.g., the Jordan-Wigner, Bravyi-Kitaev, and other transformations. 
Once the Hamiltonian is represented by a set of qubit operations, they can be executed using VQE given a suitable trial circuit on a real quantum computing hardware or a software simulator.
In order to orchestrate many such experiments and run them on different backend systems (both quantum simulator and real quantum computer), the QPFAS workflow is developed, which is described in the following.

\subsection{Workflow for scalable molecular simulation}
\label{sec:workflow_for_scalable_molecular_simulation}
\begin{figure}[htb!]
  \centering
  \includegraphics[width=0.9\textwidth]{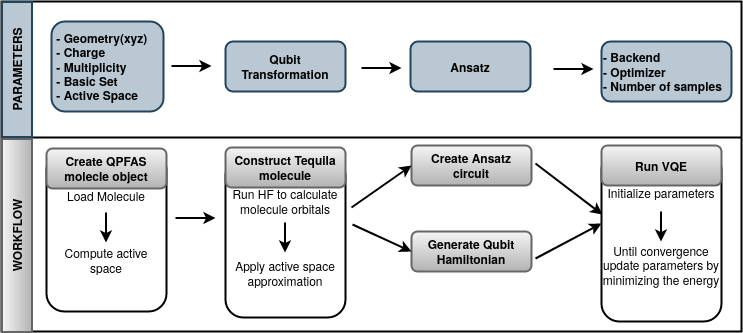}
  \caption{Scheme of the QPFAS calculation workflow. The parameters to be used in each stage are detailed in the upper part of the figure. The workflow steps are represented in boxes in the lower part of the figure. }
  \label{fig:workflow_steps}
\end{figure}

QPFAS is a computational orchestration workflow that is built around the VQE algorithm, and can be executed on gate model quantum computing hardware as well as software simulators. 
QPFAS is integrated and tested with multiple quantum simulators. 
The main simulators results presented in this paper are from the QULACS \cite{QULACS} simulator.
The results of the real quantum computer are obtained using the IonQ computer based on trapped ion technology \cite{ionq}. 
Figure~\ref{fig:workflow_steps} outlines the QPFAS workflow.
The top row shows the parameters that are to be specified at each step, while different modules are described in the second row of the figure. 
The workflow employs the Python package Tequila \cite{kottmann2021tequila} to create the quantum circuits and implement the VQE algorithm.

In the first step of the QPFAS workflow, a molecule object was instantiated to store the molecule's geometry, multiplicity, charge, basis set, and active space. 
This step provides the information required to perform classical chemistry calculations and fully specifies the chemistry problem. 
Next, the molecule is transformed in \texttt{Tequila} format\cite{kottmann2021tequila} specifying the qubit transformation to be used. 
Both objects (model formats) are used in step three to create the ansatz circuit and the qubit Hamiltonian. 
After creating these two, it is possible to run VQE using a classic optimizer in the last step of the workflow, where the number of samples and the backend (quantum computing software simulator or hardware) need to be also specified as parameters.

The QPFAS workflow yields ground state energies of molecules. 
These VQE-derived energies are benchmarked against those from classical methods such as Coupled Cluster Single-Double (CCSD) or Full Configuration Interaction (FCI) with Tequila. 
Here, it is important to mention that certain choices of active space are not compatible with Tequila's classical solver, so in these cases the Python-based Simulations of Chemistry Framework (PySCF) \cite{sun2018pyscf} is used for the classical calculations.

\subsubsection{VQE approaches in QPFAS}
\label{sec:vqe_approaches}
\begin{figure}[!htb]
  \centering
  \begin{subfigure}[b]{0.4\textwidth}
    \centering
    \includegraphics[width=0.95\linewidth]{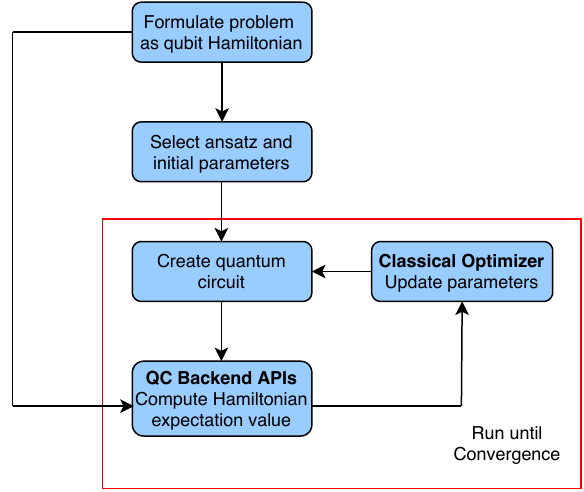}
    \caption{}
    \label{fig:workflow_vqe}
  \end{subfigure}
  \begin{subfigure}[b]{0.4\textwidth}
    \centering
    \includegraphics[width=0.9\linewidth]{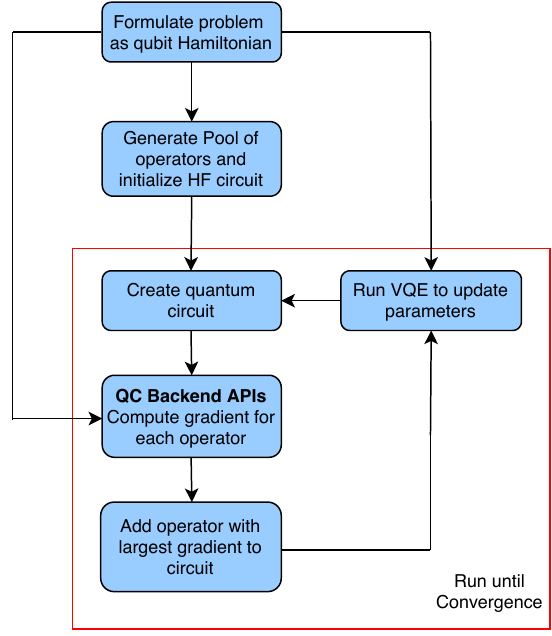}
    \caption{}
    \label{fig:workflow_vqe_adapt}
  \end{subfigure}
  \begin{subfigure}[b]{0.15\textwidth}
    \centering
    \includegraphics[width=0.9\linewidth]{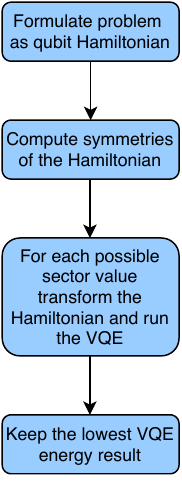}
    \caption{}
    \label{fig:workflow_vqe_tapering}
  \end{subfigure}
  \caption{ VQE enabled in PQFAS workflow: (a) Standard VQE, (b) ADAPT-VQE which grows one operator at a time in a manner set by the molecule being simulated, (c) VQE with tapering to find symmetries in the qubit Hamiltonian. Standard and ADAPT-VQE run a self-consistent loop placed inside the red box in the figure. }
  \label{fig:workflow_vqe_variations}
\end{figure}

 VQE \cite{peruzzo2014variational} is a hybrid classical-quantum algorithm that is suited for noisy intermediate-scale quantum (NISQ) computing devices.
It is an approximate method to solve optimization problems, more specifically the minimum eigenvalue problem. 
Due to the large resource requirements of other quantum algorithms, such as quantum phase estimation,\cite{Goings2022-xb} for computational chemistry problems, VQE is likely to remain the \textit{de facto} method in this field for foreseeable future. 
 In this work, in addition to the standard VQE, ADAPT-VQE and VQE by tapering are also considered as they have certain advantages for computational chemistry problems.
 The standard VQE and its variations are illustrated in Figure~\ref{fig:workflow_vqe_variations}, and desctibed in the following.

 \paragraph{Standard VQE algorithm} (Figure~\ref{fig:workflow_vqe}) 
   VQE is based on the \emph{variational principle}, which states that the smallest valid energy solution for a state will always be an upper bound to the energy of the ground state. 
   In the VQE algorithm, the molecular wave function is encoded as a parameterized circuit called the ansatz, $U(\theta)$.
   A Hamiltonian, $H$, is transformed from fermionic operators to qubit operators. 
   To find the ground-state energy, the parameters are iteratively updated to minimize the expectation value of the Hamiltonian \eqref{eqn:ground_state}, until convergence is reached.
    \begin{equation}\label{eqn:ground_state}
    E_\text{ground-state} = \min \langle 0| U(\theta)^\dagger H U(\theta) |0\rangle
    \end{equation}
    where $|0\rangle$ is the reference state the circuit acts upon usually the Hartree-Fock ground state of the system. 
    To generate the molecular Hamiltonian, the geometry of the molecule and the basis set to be used must be specified, and the necessary integrals to be computed. 
    The geometry of the molecules is stored in an \texttt{xyz} file that contains the specification and coordinates of each atom. 
    The second quantized Hamiltonian can then be written as:
    \begin{equation}\label{eqn:second_quan}
      H = h_0 + \sum_{ij} h_{ij} a_i^\dagger a_j + \sum_{ijkl} h_{ijkl} a_i^\dagger a_j^\dagger a_k a_l
    \end{equation}
    where $h_0$ is a constant, $h_{ij}$ and $h_{ijkl}$ are the single and two-body integrals respectively, and $a_i^\dagger$ is the creation operator for orbital $i$ and $a$ is the annihilation operator for orbital $i$. 
    To implement the Hamiltonian on a quantum computing hardware or software simulator, the fermionic Hamiltonian \eqref{eqn:second_quan}, should be mapped to a qubit Hamiltonian.
    This is done with both Jordan-Wigner and Bravyi-Kitaev transformations in the workflow.
    
  \paragraph{ADAPT-VQE algorithm} (Figure~\ref{fig:workflow_vqe_adapt})
    While being based on the standard VQE algorithm, the ADAPT-VQE uses an adaptive ansatz as introduced in \cite{grimsley2019adaptive}. 
    A single operator is added to the ansatz (which initially is the identity circuit) from a pool of candidate operators.
    Operators are added until convergence is reached. 
    Convergence is deemed as the gradient of energy with respect to the pool of operators. 
    Each time an operator is added to the ansatz, the VQE algorithm is run to find the optimal parameters.
    
  \paragraph{VQE algorithm with tapering} (Figure~\ref{fig:workflow_vqe_tapering})
    This variation uses the tapering algorithm \cite{bravyi2017tapering} to remove qubits before running the VQE. 
    The tapering algorithm is based on finding symmetries in the qubit Hamiltonian, where symmetry is defined by terms that commute with the Hamiltonian. 
    For each independent symmetry found, one qubit can be removed through a transformation of the Hamiltonian. 
    While this method reduces the number of qubits used, there are two downsides to be considered. 
    In this VQE variation the transformation of the Hamiltonian does not preserve the number of particles and thus only the hardware ansatz can be used as it is the only ansatz that does not conserve particle number. 
    Additionally, when removing a qubit, it is essential to know what vector to project it into ($0$ or $1$). Thus, if $n$ qubits are removed, then the VQE must be run $2^n$ times, resulting in the lowest energy value being the final result for the ground state.


\subsubsection{Ansatz calculation methods in QPFAS}  
\label{sec:ansatz_calculation_methods_in_qpfas}


In order to run VQE, it is necessary to define an ansatz.
There are generally two paradigms for VQE ansatze design: a) chemistry-inspired ansatze which are designed to capture the form of the molecular electronic wave function, and b) hardware efficient ansatze (HEAs) which attempt to offer flexible and expressive circuits that are shallower and more efficient to run on noisy hardware.
Early demonstrations of VQE on quantum hardware utilized the unitary coupled cluster with singles and doubles (UCCSD) ansatz \cite{peruzzo2014variational, OMalley2016-ph, Colless2018-hi, McCaskey2019-gk, Nam2020-ct}. 
Despite its success in treating strongly correlated systems, UCCSD becomes impractical on NISQ quantum computers due to its rapid increase in entangling gates. 
Consequently, hardware efficient ansatze (HEAs) have gained attention for requiring significantly shallower circuits. 
While HEAs have been used to simulate various molecules on superconducting quantum computers \cite{kandala2017hardware, Kandala2019-rz, Gao2019-pp, Rice2021-mn, Gao2021-zt}, HEAs struggle to deal with noise and the so-called barren plateaus in optimization \cite{McClean2018-ku}. 
Moreover, they too are subject to the measurement bottleneck \cite{Gonthier2020-rh}.
To maximize efficiency while maintaining chemical fidelity, we use a modified ansatz derived from the UCCSD ansatz: the unitary pair CCD (pUCCD) ansatz \cite{Elfving2020-xf, Elfving2020-zx}. 

In total, the QPFAS workflow allows for the use of five ansatz design methods.
Three of them are chemistry-inspired ansatze based on Unitary Coupled-Cluster Single and Double excitations and two are hardware efficient ansatze.
These five ansatze well represent the current state of the art of VQE for chemistry problems at the time this project.
The available ansatze in QPFAS, including pUCCD, are explained in the following.
\begin{itemize}
    \item \textbf{UCCSD and k-UpCCGSD}: These two ansatze are chemistry inspired. 
    In UCCSD ansatz only single and double excitations are considered, which preserve the Hartree-Fock ground state. 
    It is the original ansatz proposed for VQE in chemistry in 2014 \cite{peruzzo2014variational}. 
    For k-UpCCGSD ansatz all single excitations are taken into account (even ones that destroy the Hartree Fock ground-state), but only double excitations that move a pair of electrons from one spatial orbital to another are allowed. 
    This is a more efficient approximation to UCCSD, which has never been tested on a real quantum computer -- simulation only.
    Both these ansatze are of the form, $U=e^{T-T^\dagger}$, where $T$ is an operator generating excitations, and contain a \textit{depth} parameter. In UCCSD this corresponds to the number of trotter steps, while in k-UpCCGSD the depth ($k$) is the number of times the set of excitations are repeated (with different angles each time).
    
    \item \textbf{pUCCD}: This is also a chemistry inspired ansatz derived from the UCCSD ansatz, but only contains double excitations \cite{Elfving2020-xf, Elfving2020-zx}. 
    The pUCCD ansatz has several appealing characteristics for use on near-term hardware \cite{OBrien2022-aa}.
    First, it allows for a hard-core boson representation and requires half the number of qubits compared to other UCC ansatze. 
    Second, at most three circuits are needed for energy expectation value computation, eliminating the measurement bottleneck for this ansatz (measurement is $\mathcal{O}(1)$). 
    Finally, it has entangling gate requirements that goes as $\mathcal{O}(N^2)$ and entangling gate depth that goes as $\mathcal{O}(N)$ \cite{Zhao2023-um}.

    \item \textbf{``hardware'' and ``hardware conserving''}: These two ansatze consider quantum computing hardware. \textit{hardware} ansatz \cite{kandala2017hardware} and \textit{hardware conserving} ansatz \cite{barkoutsos2018quantum} both consist of repeating layers of rotation gates (acting on individual qubits) followed by an entangling layer.
    In the case of \textit{hardware} ansatz the circuit acts on the null state (all qubits set to zero) and the entangling layer is a set of CNOT gates. 
    The \textit{hardware conserving} ansatz acts on the Hartree-Fock ground state and the entangling layer is designed to conserve the number of particles.
\end{itemize}

\subsubsection{Orbital reduction methods in QPFAS}
\label{sec:orbital_reduction_methods_in_qpfas}
In quantum computer the number of qubits required to run a circuit is directly related with the number of spin orbitals of the problem. Hence, using a subset of the orbitals instead of the entire set requires a significantly fewer qubits for implementation. 
The QPFAS workflow supports two automatic methods to find such a subset of orbitals: \textit{frozen core approximation} and \textit{Natural Orbital Occupation Numbers (NOONs)}.
It is also possible to provide a manual input. 
These are referred to as orbital reduction methods.
For background, the orbitals are categorized as \textit{occupied}, \textit{frozen}, or \textit{virtual}. 
Usually, the inner occupied orbitals are called \textit{core orbitals}. 
These can be assumed to be frozen, i.e., not participant in the excitations. 
The unoccupied orbitals are called \textit{virtual}. 
The remaining occupied orbitals and the virtual orbitals are referred to as the \textit{active set}. 
Thus the Hamiltonian is expressed as
\begin{equation}\label{eqn:hamiltonian}
H = \sum_{i\in F}h_{ii}  + \sum_{ij \in A} \left( h_{ij}  + \sum_{k \in F} (h_{kijk} - h_{kikj}+h_{ikkj}-h_{ikjk})  \right) c_i^\dagger c_j
+ \sum_{ijkl \in A}h_{ijkl} c_i^\dagger c_j^\dagger c_k c_l 
\end{equation}
where $F$ is the set of frozen orbitals and $A$ is the active set.

The \textit{frozen core approximation} treats the lowest energy orbitals as being occupied, assuming that the core electrons are not correlated. 
The number of orbitals to freeze is determined based on the atomic species of the molecule\footnote{ORCA Input Library, Frozen core calculations. \url{https://sites.google.com/site/orcainputlibrary/frozen-core-calculations}}.   
For \textit{Natural Orbital Occupation Numbers (NOONs)}, the single particle density matrix, $\rho(x, y)$ is defined by the set of basis functions $\{\phi_i(x)\}$, given by
\begin{equation}\label{eqn:density}
     \rho(x, y) = \sum_{ij} \phi_i(x) \gamma_{ij} \phi_j(y)^*.
\end{equation}    
Since $\gamma_{ij}$ is a hermitian matrix we can diagonalize it, where the resulting diagonal matrix contains the NOONs. 
For Hartree-Fock, the NOONs are either zero or one. 
By computing the NOONs with a higher-order method (such as MP2) we get a non-trivial result. 
If the NOONs for a given orbital is close to zero (or two) within a specified tolerance, then the orbital is considered to be unoccupied (doubly occupied).

\section{Experiments and results}
\label{sec:experiments_and_results}

Due to the limitations of quantum computing simulators and hardware,  we study the bond-breaking for smaller PFAS-like molecules suach as \ch{CH3F}.
This lays the groundwork to study larger molecules with C-F bonds in the future. 
Carrying out these experiments with QPFAS workflow allows integrating HPC models and IonQ real quantum hardware. 
Another objective is to explore the current computational limits for quantum chemistry calculations using an end-to-end hybrid workflow.
Hence, the following experiments seek to explore three main characteristics: 
\begin{itemize}
    \item \textbf{Scalability} of the problem in terms of the size of the molecules;
    \item \textbf{Functionality of the QPFAS workflow} with the largest bond-breaking experiment that can fit into our HPC or a real quantum computer;
    \item \textbf{VQE accuracy} and its performance under a wide range of parameterizations. 
\end{itemize}
The experiments executed on the HPC simulator serve to refine the parameters for running a bond-breaking experiments in the real quantum computer.
The experiment and the results are divided in two subsection: first, calculations in QULACS simulator on ICHEC HPC cluster, and second, calculations in the IonQ quantum computer.

\subsection{Experiments using QULACS software simulator}
\label{sec:experiments_using_quantum_computing_software_emulation}
The first part of the experimentation is focused on gaining knowledge on scalability and parameter optimization running calculations on a group of eleven molecules with increasing size. The second part is dedicated to explore energy dissociation only on \ch{F2} and \ch{CH3F}. 
Then the bond stretching calculations of these two molecules are executed, as they are the biggest molecules we can execute in the ICHEC's HPC in a reasonable amount of time (less than 24 hours). \ch{F2} molecule is selected because it is one of the largest with an F-F bond that we could run on the quantum emulator.
\ch{CH3F} is the largest molecule that can be executed with a C-F bond in the quantum emulator.

\subsubsection{Scalability and parameterization study}
\label{sec:results_from_quantum_computing_software_emulation}
VQE is a heuristic quantum algorithm that scales exponentially when the number of variables of the problem increases.
Hence, there is a practical hard limit on the problem size when executed on a classical quantum simulator.
To avoid unbounded runtime of the problems on the simulator, a timeout limit of 24 hours is set in the executions. 
All molecules/runs that exceed this timeout are excluded from the analysis of the results.

\begin{figure}[!htb]
	\centering
	\includegraphics[width=\textwidth]{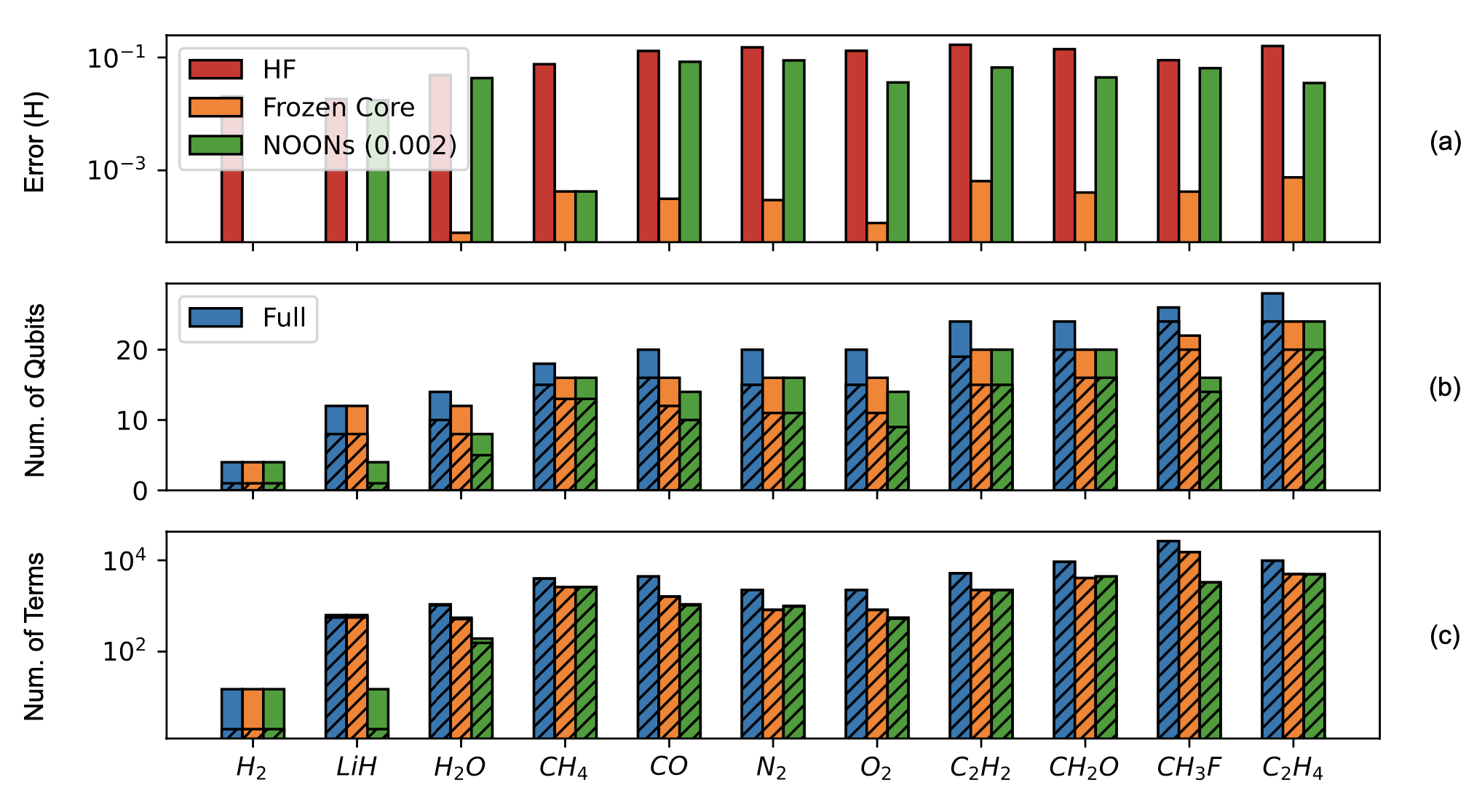}
	\caption{Comparison of three active space methods, HF in red, Frozen Core in orange and NOONs in green for (a) panel. Middle (b) and bottom (c) panels correspond to number of qubits and number of Hamiltonian terms as function of molecule size respectively. Blue bars refer to \textit{full} set of orbitals included in the calculation. The hatching refers to tapering on top of the active space
approximations.}
	\label{fig:comparison_of_active_space_methods}
\end{figure}

First study is to explore the active space selection methods as a function of the size of the molecules, as presented in Figure~\ref{fig:comparison_of_active_space_methods}. 
This shows the number of qubits needed to represent the problem and the number of terms in the Hamiltonian.
The QULACS quantum circuit simulator runs on the Kay cluster\footnote{ICHEC HPC cluster, Kay. \url{https://www.ichec.ie/about/infrastructure/kay}} typically on a single high-memory node. 
Hartree-Fock, Frozen Core and NOONs are selected as active space methods to to compare their error levels as a function of molecule size (Figure~\ref{fig:comparison_of_active_space_methods} (a)).
For the NOONs method, the upper threshold parameter of 0.002 is selected based on experiments yielding errors within the acceptable range.

Figure~\ref{fig:comparison_of_active_space_methods} shows how the error (compared against classically-computed CCSD as a baseline), number of qubits and number of Hamiltonian terms vary with choice of active space. 
In the top panel (a) compares the error levels to CCSD results as a function of molecule size.
HF and NOONs errors scale with number of electrons while Frozen Core stay stable. 
The middle panel (b) shows the number of qubits reduced by using the frozen core and NOONs approximations. 
The bottom panel (c) show the impact on the number of terms in the Hamiltonian, \ch{CH3F} being the most impacted. 
The hatching refers to tapering on top of the active space approximations.

\begin{figure}[!htb]
\centering
\includegraphics[width=\textwidth]{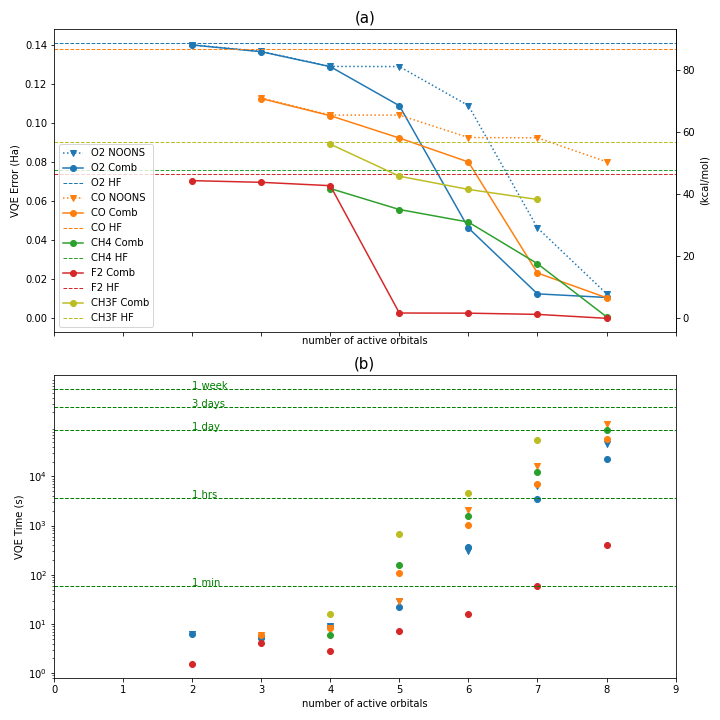}
\caption{Scaling up of the problem in terms of \textbf{number of orbitals}. Upper panel (a) shows VQE error for \ch{O2}, \ch{CO}, \ch{CH4}, \ch{F2} and \ch{CH3F} as a function of the number of active orbitals. VQE error is expressed in Hartrees (Ha) in left vertical axis and in kcal/mol in right vertical axe. Three reduction orbital methods where applied , NOONs, Comb and HF . Lower panel (b) is a plot of VQE time (logarithm scale) to show the scaling up in terms of active orbitals for the same group of molecules than upper panel}
\label{fig:qubit_number_active_orbitals}
\end{figure}

For a group of selected molecules (only five to simplify the visualization), Figure \ref{fig:qubit_number_active_orbitals} illustrates the VQE error in upper panel (a) and the VQE runtime in bottom panel (b) as a function of the number of active orbitals. 
For these molecules we run experiments using three orbital reduction methods NOONs, Comb and Hartree-Fock. 
In average, the VQE time scales up when the number of active orbitals and the molecule size increase, as expected. 
The results for all molecules tested is available on the QPFAS GitHub page\footnote{Qubit number scaling \url{https://github.com/ICHEC/QPFAS/blob/0.4.0-beta/py/examples/circuit_requirements/all_qubit_number_experiments.png}}.

\begin{figure}[!htb]
\centering
\includegraphics[width=\textwidth]{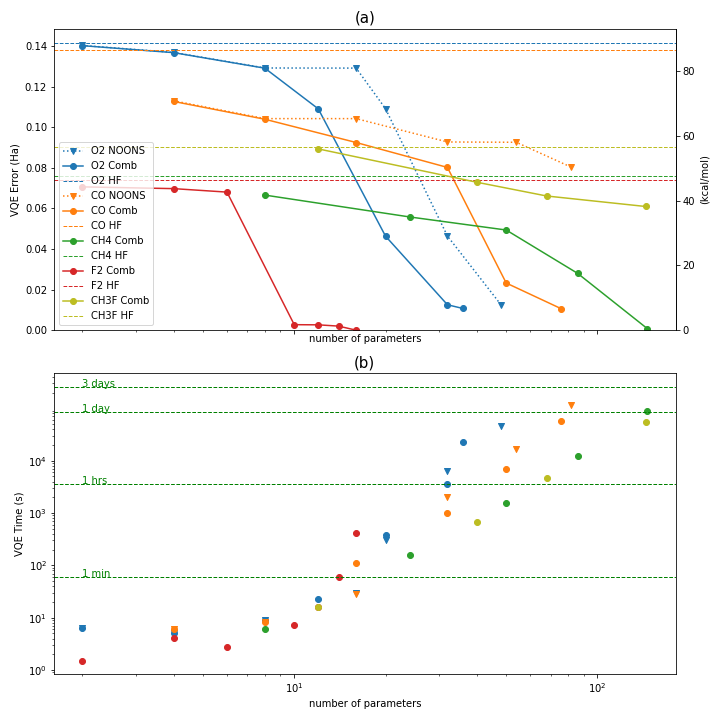}
\caption{Scaling up of the problem in terms of \textbf{ VQE ansatz parameters}. Upper panel (a) shows VQE error for \ch{O2}, \ch{CO}, \ch{CH4}, \ch{F2} and \ch{CH3F} as a function of VQE ansatz parameters. VQE error is expressed in Hartree(Ha) in left vertical axe and in kcal/mol in right vertical axe. Three reduction orbital methods where applied NOONs, Comb and HF. Lower panel (b) shows VQE time (logarithm scale) in terms of active orbitals for the same group of molecules than upper panel. The error levels can be systematically lowered by increasing the complexity of the VQE ansatz, which translates to increased computation times/resources.}
\label{fig:parameter_number_experiments}
\end{figure}

For the same group of selected molecules as above, Figure \ref{fig:parameter_number_experiments} shows how the problem scales in terms of the number of parameters of the VQE ansatz for increasing active orbitals and molecule size. 
The plot shows the ramp up in calculation time by increasing the VQE ansatz parameters and the size of the molecules, as expected.
It also confirms that error levels can be systematically lowered by increasing the complexity of the VQE ansatz, which translates to increased computation times/resources.
The results for all molecules tested is available on the QPFAS GitHub page. \footnote{Number of parameters scaling \url{https://github.com/ICHEC/QPFAS/blob/0.4.0-beta/py/examples/circuit_requirements/all_parameter_number_experiments.png}}.

\paragraph{Results for parameter optimisation}

Parameter optimization is a set of tests using \ch{H2} and \ch{HF} molecules to find best parameters of the QPFAS workflow for larger simulations such as \ch{CH3F} molecule. 
The studied parameters include a list of optimizers (refer to the figure in Github for details), the qubit transformations Jordan-Wigner and Bravyi-Kitaev, a the range of five ansatze, and different sample sizes.
We concluded that hardware ansatze are less performant than chemistry inspired ansatze. 
In addition, the results show that the use of one transformation or the other does not privilege any optimizer. 
However,  \textit{Powell} is the optimizer that presents the smallest error for kUpCCGSD, tapering and UCCSD in both transformations. 
We also investigate the number of samples need to successfully run the VQE algorithm. This study allow us to identify the number of samples against final VQE energy for the \ch{H2} molecule. 
 We run the workflow a total of five times to measure the variance of the results. 
 It can be inferred from the results (refer to the figure in Github for details) that more than 10,000 samples must be used to achieve an error less than $0.001$ Ha.
The results also indicate that the BFGS optimizer performs better than COBYLA in small sample sizes.

\paragraph{Results for bond stretching}

Here we present the results on how well the VQE can describe the energy profile of a molecule when bonds are stretched and broken. 
The previous experiments helped to select a set of optimal parameters to run the bond-breaking experiment.
These experiment are done for \ch{F2} and \ch{CH3F} molecules as they can give us more information about the breakdown in PFAS substances regarding C-F and F-F bonds.
Standard VQE with UCCSD ansatz is used for \ch{F2}, and ADAPT-VQE with UCCSD is for \ch{CH3F}.
ADAPT-VQE is the most performant ansatz at the time this computation is run. 
To map spin operators onto fermionic operators we use Jordan-Wigner transformation for both molecules.

\begin{figure}[!htb]
\centering
\includegraphics[width=\textwidth]{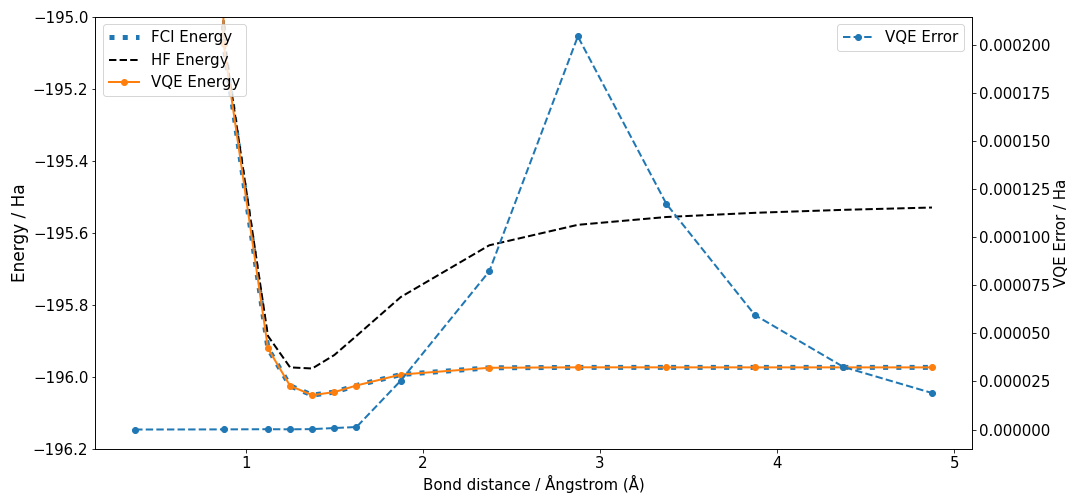}
\caption{Energy as a function of bond stretching for \ch{F2} molecule. Three methods are calculated, blue dotted line corresponds to FCI, dashed black line to HF and VQE standard UCCSD with Jordan-Wigner transformation in orange dot points and full line. FCI and HF are superposed. Left vertical axe corresponds to energy values expressed in Hartree(Ha), right vertical axe corresponds to error of VQE calculation expressed in Ha units, the values of computed VQE error are plotted with dashed blue line and dot points. The bond distance in horizontal axe is in Angstrom(A).}
\label{fig:f2_using_uccsd_and_bond_stretching}
\end{figure}

Figure \ref{fig:f2_using_uccsd_and_bond_stretching} shows the dissociation curves of the \ch{F2} molecule as a function of bond distance. 
The three curves of energy follow the same trend for lower bond distances, however HF shows higher energy values when the bond starts to be dissociated. 
FCI produces the same results as HF. 
The right vertical axe of the figure shows the VQE error, which reaches its maximum value when the molecule is in the dissociation stage.  
The error is quite small during all the calculation steps.

\begin{figure}[!htb]
\centering
\includegraphics[width=\textwidth]{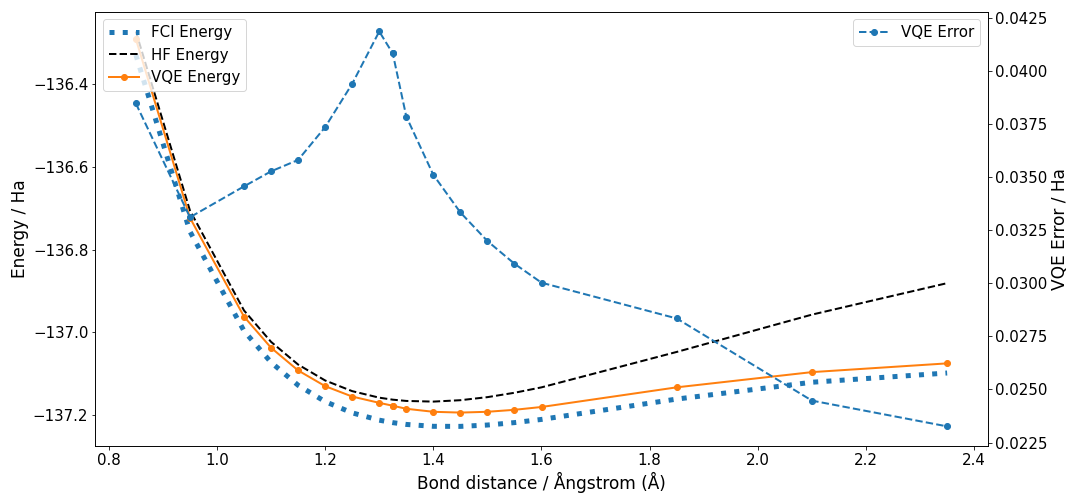}
\caption{\ch{CH3F} using ADAPT-VQE, JW transformer, UCCSD and bond stretching . Energy as a function of bond stretching for \ch{CH3F} molecule. Three methods are calculated, blue dotted line corresponds to FCI, dashed black line to HF and ADAPT-VQE UCCSD with Jordan-Wigner transformation in orange dot points and full line. Left vertical axe corresponds to energy values expressed in Hartree (Ha), right vertical axe corresponds to error of VQE calculation expressed in Ha units, the values of computed VQE error are plotted with dashed blue line and dot points. The bond distance in horizontal axe is in Angstrom(A).}
\label{fig:ch3f_using_adapt_vqe_uccsd_and_bond_stretching}
\end{figure}

Figure \ref{fig:ch3f_using_adapt_vqe_uccsd_and_bond_stretching} shows the bond dissociation curve for \ch{CH3F} by stretching the C-F bond. The energy from HF is above the VQE result during the entire dissociation process, which becomes much more significant when the C-F bond is completely broken. 
However, FCI energy remains below the energy values obtained with ADAPT-VQE. 
For \ch{CH3F} we observe that the VQE error is maximum at the ground state energy unlike what is observed for \ch{F2}. 
Even if the error is quite small during all the calculations in both molecules, we can observe that the error is two orders of magnitude smaller for \ch{F2}.
The results show that VQE can capture correlations and provide accuracy beyond HF, especially for a bond-breaking experiment. 
Considering that FCI is not an scalable approach by any means, this proves VQE as a valid alternative for post-HF quantum chemistry calculation.
This is specially important due to the suitability of VQE for quantum computing technology, as explored in the next subsection.

\subsection{Experiments using IonQ's Aria trapped-ion quantum computer}
\label{sec:computational_chemistry_challenges_in_pfas}

A key objective of this work is to run experiments on currently available quantum processing unit (QPU) devices, while looking ahead to advances in quantum hardware that will enable the simulation of larger  PFAS-like molecules. 
The IonQ Aria quantum computer is integrated in the QPFAS workflow as the quantum hardware where the end-to-end process can be executed. 
The integration allows us to run shallower circuits with higher fidelity, as a result of the accuracy and all-to-all connectivity of the trapped-ion architecture.
Hence, the results are less affected by the noise in quantum computing devices. 
%
Given the capabilities of the current quantum hardware, we tighten our focus to evaluating the scalability of the QPFAS workflow running the C-F bond dissociation on a small PFAS-mimic molecule (\ch{CH3F}). 
For this experiment, we do not consider the convergence behaviour of the pUCCD ansatz and instead obtain the classically-optimized parameters for the VQE algorithm. 
This is a common approach on VQE studies both on trapped-ion\cite{Khan2022-cu, Zhao2023-um} and superconducting devices \cite{OBrien2022-aa}.
VQE’s flexibility in circuit design allows fine tuning based on quantum gate fidelity, number of qubits, and desired accuracy, making it well-suited for the NISQ era.
However, running shallower circuits in VQE comes with its costs: predicted energy remains approximate, and many more measurements are required.

At the time the simulations were performed (late 2022) IonQ had two publicly available quantum computers to run the experiments proposed in this work; an 11-qubit trapped-ion QPU (Harmony) and a 25-qubit trapped-ion QPU named Aria. 
As Aria's gate fidelities are significantly higher than the previous generation (Harmony) and it has sufficient qubit counts to simulate PFAS-like molecules, we decided to perform the experiments on the Aria QPU in order to achieve one of the largest quantum chemistry simulations on real quantum hardware at the time. 

As outlined in previous sections, the C-F bond is one of the strongest bonds in chemistry. Thus, we evaluate PFAS-like molecules to break the C-F bond of them. 
The \ch{CH3F} molecule is chosen as the largest molecule for the experiments. 
Hence, we carry out the next two experiments on the quantum computer Aria to meet the objectives of the study:
\begin{enumerate}

    \item \textbf{\ch{CH3F} equilibrium geometry}: we choose the equilibrium geometry of the \ch{CH3F} molecule to calculate its ground state. 
    Since a sull VQE algorithm would require a lot of computational time being executed in a real quantum hardware, we perform most of the VQE iterations using IonQ's cloud statevector simulator\cite{noauthor_undated-uy} (to obtain the values of the optimized parameters of the ansatz).
    Then, we execute the final energy evaluation using the IonQ Aria quantum computer.
    \item \textbf{\ch{CH3F} stretched geometry}: we set another experiment around the C-F bond breaking of the \ch{CH3F} molecule. For this experiment, a geometry of the molecule is defined further away from the equilibrium geometry, trying to simulate how the stretch would be. 
    Similar to the above case, we execute almost all the iterations of the VQE in the IonQ statevector simulator, then the final energy evaluation is done using IonQ Aria.
\end{enumerate}


These three experiments are performed under the same parameterization of the QPFAS workflow, which is:
\begin{itemize}
    \item \textbf{Basis set}: STO-3G basis set is used to create the QPFAS molecule.
    \item \textbf{Active space}: the \textit{frozen core} approximation is used to create the QPFAS molecule. In case of the \ch{CH3F} molecule, it reduces the number of qubits by 2, leaving the number of active orbitals on 11.
    \item \textbf{Transformation}: Jordan-Wigner transformation is used to create the qubit Hamiltonian.
    \item \textbf{Ansatz}: the pUCCD ansatz is used as it is an efficient shallow ansatz, and has shown high-performance in recent quantum experiments both at Google\cite{OBrien2022-aa} and IonQ\cite{Zhao2023-um}. 
    In the pUCCD ansatz, the \ch{CH3F} molecule has 11 qubits and 28 variational parameters after freezing the core electrons, with no more than 253 single qubit gates and 56 entangling gates.
\end{itemize}

\subsubsection{Results from IonQ quantum computer and error mitigation}
\label{sec:results_from_ionq_quantum_computer}



\begin{table}[!htb]
\caption{\ch{CH3F} energy of equilibrium geometry}
\centering
\begin{tabular}{lrrrrrr}
\hline
                               & \multicolumn{5}{c}{threshold + symmetry post-selection}        & \multicolumn{1}{l}{} \\ \hline
Raw energy (Hartree)           & 0.0\%      & 0.5\%      & 1.0\%      & 1.5\%      & 2.0\%      & exact                \\ \hline
\multicolumn{1}{r}{-135.74352} & -136.54827 & -136.76696 & -136.85525 & -137.14006 & -137.19404 & -137.19512           \\ \hline
\end{tabular}\label{tab:ch3f_equilibrium_energy}
\end{table}

\begin{table}[!htb]
\caption{\ch{CH3F} energy of stretched geometry}
\centering
\begin{tabular}{lrrrrrr}
\hline
                               & \multicolumn{5}{c}{threshold + symmetry post-selection}        & \multicolumn{1}{l}{} \\ \hline
Raw energy (Hartree)           & 0.0\%      & 0.5\%      & 1.0\%      & 1.5\%      & 2.0\%      & exact                \\ \hline
\multicolumn{1}{r}{-136.11887} & -136.68776 & -136.84109 & -136.85204 & -137.06390 & -137.06390 &   -137.11248    \\ \hline
\end{tabular}\label{tab:ch3f_stretched_energy}
\end{table}
Working in the noisy intermediate-scale quantum (NISQ) era, the results are subject to errors. 
For VQE, where the expectation value of an electronic Hamiltonian is to be estimated to high degrees of precision, noise can significantly impact the computed results \cite{peruzzo2014variational, OMalley2016-ph, Colless2018-hi, McCaskey2019-gk, Nam2020-ct, kandala2017hardware, Kandala2019-rz, Gao2019-pp, Rice2021-mn, Gao2021-zt, Zhao2023-um}. Although the purpose of this work is not to showcase error mitigation protocols, it can be seen from the raw data that the stretched and equilibrium geometries are in qualitative disagreement: the stretched geometry is \emph{lower} in energy than the equilibrium geometry, which is incorrect both theoretically and physically. 
To mitigate this discrepancy and also to show a simple path to recovering near-ideal results, we exploit two symmetries of the chemical problem to mitigate the impact of noise on the results.

The first method we use is symmetry post-selection for the $\langle Z \rangle$-basis measurements \cite{Bonet-Monroig2018-xa}. 
Given the Jordan-Wigner encoding and the hard-core bosonic approximation inherent in pUCCD, the Hamming weight of all symmetry-allowed bitstrings (output states) corresponds to the number of electron pairs. 
Any output bitstring that deviates from the ideal Hamming weight can be discarded, as it can only arise in error due to noise. Once the erroneous bitstrings are discarded, the symmetry-corrected histogram may be $L_1$-renormalized. Following this, we can see in Tables~\ref{tab:ch3f_equilibrium_energy} and \ref{tab:ch3f_stretched_energy} that from symmetry post-selection alone, a reduction in error on the order of 250-500 mHartree can be obtained. 
Despite this massive improvement in accuracy, however, symmetry post-selection is not sufficient in this case to fix the qualitative discrepancy between the energy ordering of the two \ch{CH3F} geometries.

\begin{figure}[!htb]
\centering
\includegraphics[width=\textwidth]{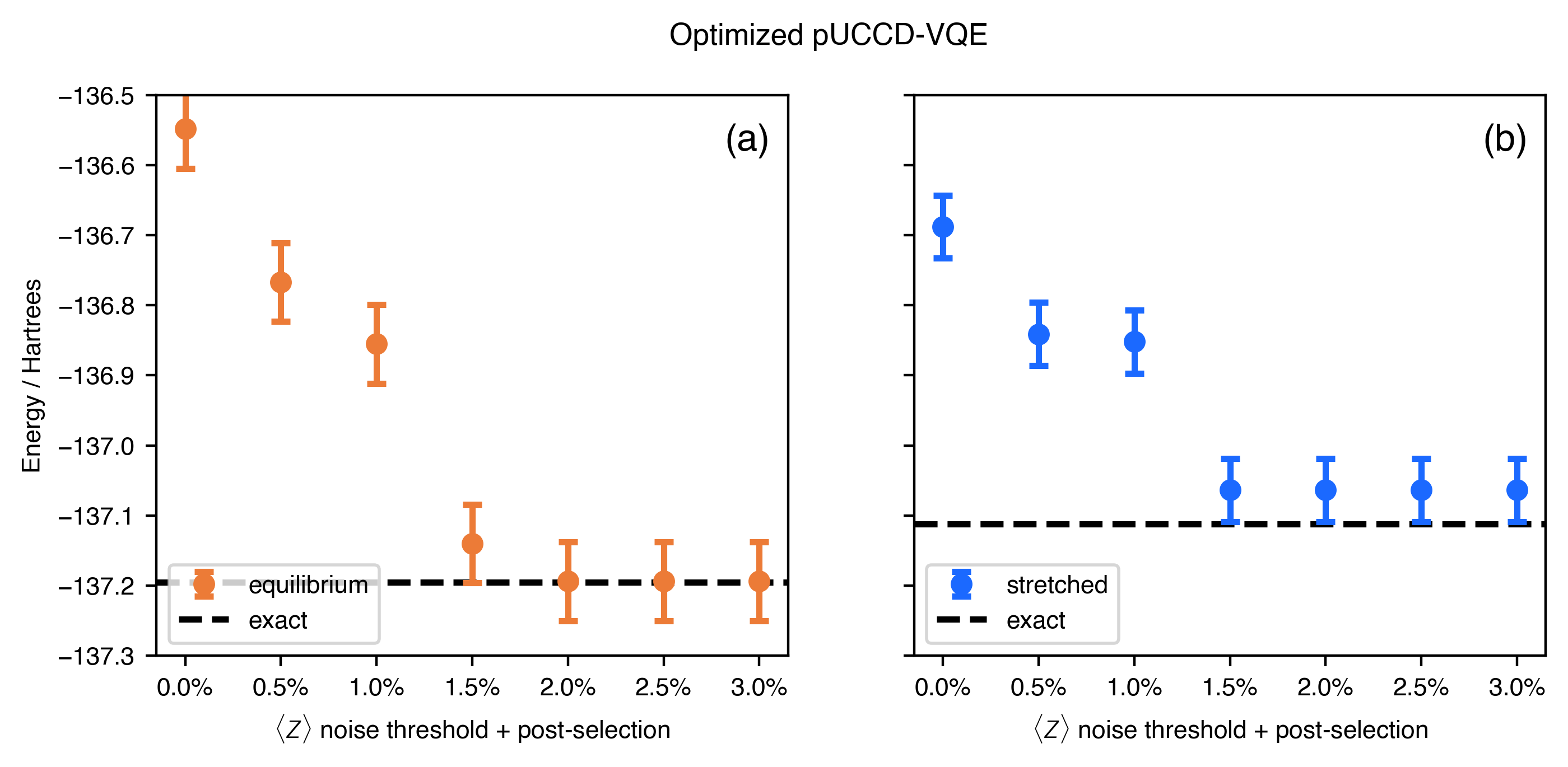}
\caption{Results of thresholding-based noise mitigation and symmetry post-selection on the $\langle Z \rangle$-basis measurements for (a) equilibrium and (b) stretched \ch{CH3F} geometries with optimized pUCCD parameters. Histogram bitstrings with probability less than the threshold value are discarded, along with bitstrings that are known to be wrong by symmetry (e.g. not particle conserving). The histograms are then $L_1$-renormalized. From the known Hartree-Fock results in Fig.~\ref{fig:hartree-fock_ch3f_noise_mitigation}, we can determine that the empirically optimal noise-floor threshold is at probabilities $\leq 1.5\%$. Error bars correspond to 2$\sigma$, where $\sigma$ is the standard error of the estimated expectation values. }
\label{fig:puccd_ch3f_noise_mitigation}
\end{figure}

\begin{figure}[!htb]
\centering
\includegraphics[width=\textwidth]{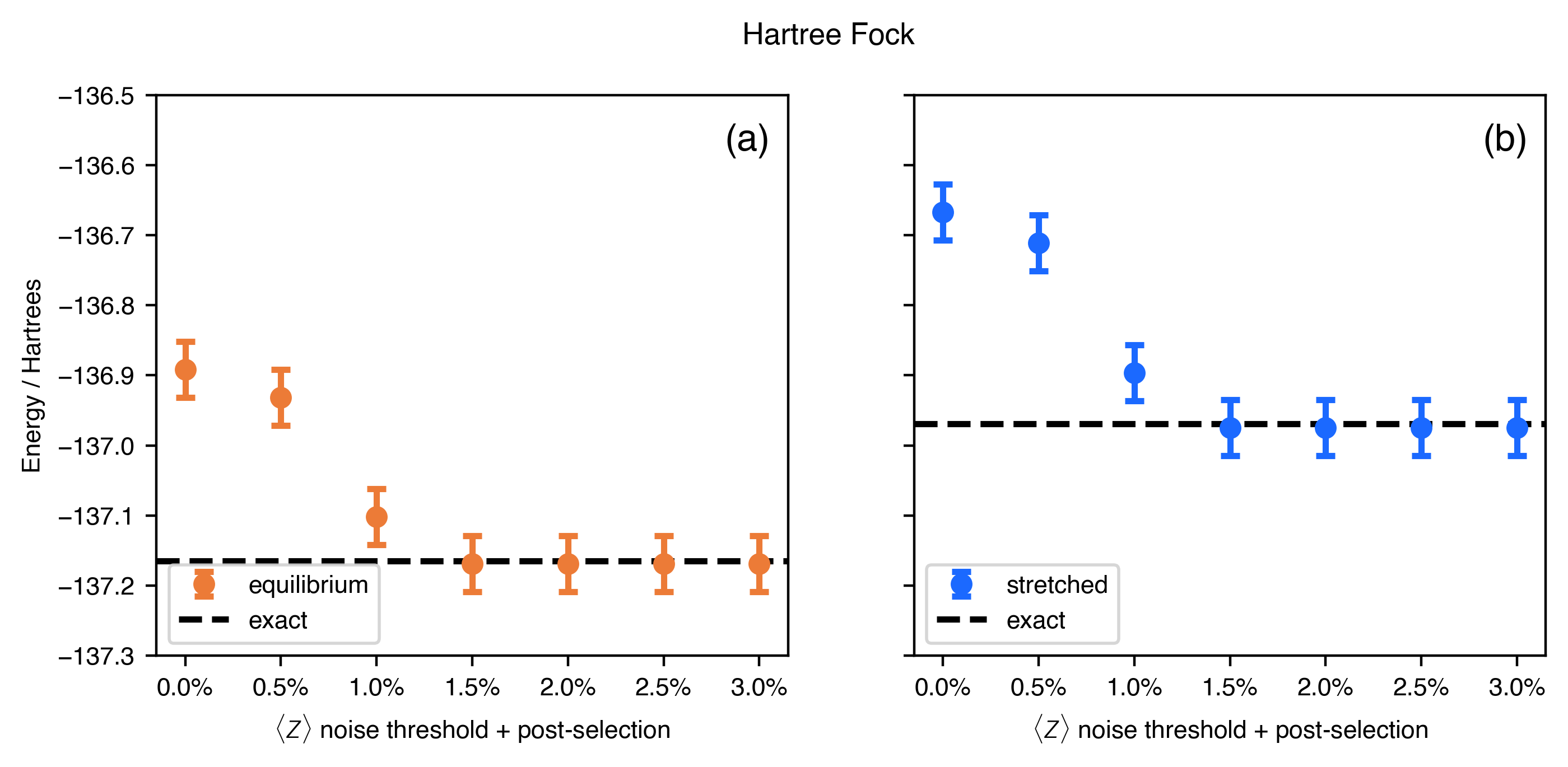}
\caption{Results of thresholding-based noise mitigation and symmetry post-selection on the $\langle Z \rangle$-basis measurements for (a) equilibrium and (b) stretched \ch{CH3F} geometries with Hartree-Fock parameters. Histogram bitstrings with probability less than the threshold value are discarded, along with bitstrings that are known to be wrong by symmetry (e.g. not particle conserving). The histograms are then 1-norm renormalized. Because the Hartree-Fock energy is efficiently and exactly known, and moreover the circuit topology is identical to the optimized pUCCD case, the Hartree-Fock hardware experiments allow us to estimate the noise floor, which can then be applied to identical circuits where the energy is not \emph{a priori} known. Error bars correspond to 2$\sigma$, where $\sigma$ is the standard error of the estimated expectation values.}
\label{fig:hartree-fock_ch3f_noise_mitigation}
\end{figure}

The second error-mitigation technique is inspired by the fact that by setting all variational parameters to zero, the Hartree-Fock energy---which is efficiently known classically---can be computed for a given (noisy) circuit. 
Moreover, in the $\langle Z \rangle$-basis, the ideal output corresponds to one and only one bit-string, corresponding to the state preparation circuit. 
Hence, by learning the ideal output for the quantum circuit, the empirical impact of noise can be estimated as the proportion of the output that deviate from the ideal \cite{Urbanek2021-aw}. 
In analogy with other signal processing techniques, such as amplitude-limiting, clipping, and threshold-based peak detection, we discard small states that cannot be reliably distinguished from noise. 
A similar approach is described in Ref ~\cite{Montanaro2021-yv}. Therefore, applying different truncation thresholds to the Hartree-Fock states (Fig. ~\ref{fig:hartree-fock_ch3f_noise_mitigation}) we see that discarding states with probability lower than 1.5\% yields a good balance between removing noise and (potentially) discarding signal. Since the circuits and hardware are the same (differing only in the value of the variational parameters) estimating the noise floor from the Hartree-Fock outputs can be applied to the optimized circuit, where the output is not in general known \emph{a priori}. 
The impact of different noise-floor thresholds on the $\langle Z \rangle$-basis can be observed in Fig.~\ref{fig:puccd_ch3f_noise_mitigation} as well as in Tables ~\ref{tab:ch3f_equilibrium_energy} and \ref{tab:ch3f_stretched_energy}. It can be seen that the empirically-determined threshold of discarding states below 1.5\% is sufficient to restore the qualitative correctness of the computed results on quantum hardware, as well as provide near-exact quantitative accuracy. 
This method of truncation applies only to the $\langle Z \rangle$-basis measurements, because the ideal output is a single bitstring. 
The $\langle X \rangle$- and $\langle Y \rangle$-basis measurements are uniformly distributed in the Hartree-Fock case, preventing this type of noise mitigation.

To further improve the results, more advanced noise mitigation techniques could be used, as this is an active and ongoing area of research \cite{Bonet-Monroig2018-xa,Montanaro2021-yv,Bennewitz2022-fo,Quek2022-ly,Kandala2019-rz,OBrien2022-aa,Weber2021-eg}. 
However, by using simple noise mitigation methods, we show here that it is possible to deliver quantum value on today's NISQ devices.

\section{Conclusions}
\label{sec:conclusions_and_actions}

Remediation of global PFAS pollution is a an active research topic that aims at finding new mechanisms to destroy these substances. 
A major pathway for PFAS destruction is breaking carbon-fluorine bonds.
Computational simulating of this process can speed-up the research towards finding optimal processes and catalysts. 
Solving the electronic Schrodinger problem is at the core of computational chemistry, which has always been a technical challenge due to its exponential complexity.
The original idea of quantum computing came from this need for simulating nature at quantum level. 
In this article, we proposed the QPFAS workflow, a hybrid quantum-classical  workflow for end-to-end chemistry simulation based on VQE algorithm.
The QPFAS workflow allows flexible definition of many experiments, including bond-breaking calculations, and orchestrates the execution of sub-tasks on both HPC-based simulators and quantum computers. 

We completed simulation experiments on a group of molecules to understand how the VQE algorithm scales in terms of molecule size, number of active orbitals, and optimization parameters. 
Two molecules that approach PFAS substances were identified, \ch{F2} and  \ch{CH3F}, with which breaking-bond simulation experiments have been carried out on a quantum computing simulator, reaching the limits of classical computing at ICHEC's HPC system.
The results on quantum simulators were consistent with the expected theoretical analysis, and helped to better understand the scalability and accuracy of the VQE algorithm. 
They provided important understanding on how to adjust the parameters and port the calculations to a real quantum computer.

Moreover, we studied the bond breaking of \ch{CH3F} molecule with IonQ's trapped-ion quantum computer.
To address the limitations of noisy quantum computing technology, we introduced two error mitigation techniques: (a) the symmetry post-selection for the ⟨Z⟩-basis measurements, and (b) correcting the error bias by calculating the classically-known Hartree-Fock energy in the noisy device.
This error mitigation approach returned near-quantitative results, with the \ch{CH3F} equilibrium energy at milli-Hartree precision with the expected result. The accuracy of the results obtained for the \ch{CH3F} molecule is unprecedented on a quantum computer for any bond-breaking experiment of a comparable size. 

The creation of the QPFAS workflow allowed the hybrid study of the bond dissociation problem with great flexibility in parameterization and orchestration of large number of parallel tasks. 
This enabled us to optimize the quantum computer resources by selecting the most efficient parameters, leading to one of the largest-ever molecules studied on computer computer.
The workflow can be used to conduct a wide range of computational chemistry experiments, and further develop the VQE algorithm for quantum computers.

\section*{Acknowledgments}
This project was supported in part by Enterprise Ireland Innovation Partnership Programme (IP/2019/0835) and EuroCC in Ireland (grant agreement no. 951732).

\bibliographystyle{IEEEtranN}  
\bibliography{references, jjgoings_qpfas}  

\end{document}